\documentclass[prl, aps, twocolumn]{revtex4} 
\usepackage{amssymb}
\usepackage{amsmath,amsthm}
\usepackage{graphicx}
\begin{document}

\title{A smooth cascade of wrinkles at the edge of a floating elastic film}
\author{Jiangshui Huang$^{1,2}$, Benny Davidovitch$^{1}$, Christian Santangelo$^{1}$, Thomas P. Russell$^2$, and Narayanan Menon*$^1$}

\affiliation{%
$^{1}$Department of Physics, University of Massachusetts, Amherst,
Massachusetts 01003 }
\affiliation{%
$^{2}$Polymer Science and Engineering Department, University of
Massachusetts, Amherst, Massachusetts 01003}

\date{\today}

\begin{abstract}

\end{abstract}


\maketitle

\textbf{The mechanism by which a patterned state accommodates the
breaking of translational symmetry by a phase boundary or a sample
wall has been addressed in the context of Landau branching in type-I
superconductors \cite{Huebener}, refinement of magnetic domains
\cite{HubertSchafer}, and compressed elastic sheets
\cite{PomeauRica}. We explore this issue by studying an ultrathin
polymer sheet floating on the surface of a fluid, decorated with a
pattern of parallel wrinkles. At the edge of the sheet, this
corrugated profile meets the fluid meniscus. Rather than branching
of wrinkles into generations of ever-smaller sharp folds
\cite{PomeauRica}, we discover a smooth cascade in which the coarse
pattern in the bulk is matched to fine structure at the edge by the
continuous introduction of discrete, higher wavenumber Fourier
modes. The observed multiscale morphology is controlled by a
dimensionless parameter that quantifies the relative strength of the
edge forces and the rigidity of the bulk pattern.}

When a thin rectangular sheet floating on the surface of a pool of
liquid is compressed along two opposing edges, it forms a pattern of
wrinkles parallel to these edges. Unlike the Euler buckling of an
unsupported piece of a paper, where the largest possible wavelength
is selected, the wrinkles form at a wavelength $\lambda\!\ll\! W$,
the width of the rectangle in the direction of the compression.  Two
principles are essential to understanding the amplitude and
wavelength of this pattern: first, a thin sheet can, to a first
approximation, be treated as inextensible, so that the length of a
line in the compression direction is preserved. Consequently, the
wavelength and amplitude of the wrinkles are related. Second, the
bending energy of the sheet favours long wavelengths (and therefore
large amplitudes) whereas the gravitational energy of the liquid
subphase favours small amplitudes (and hence small wavelengths).
Thus, the wavelength is selected by a compromise \cite{Cerda03}
between the bending energy of the sheet and gravitational energy (or
more generally, any mechanism by which the subphase resists
distortion).

\begin{figure}
\includegraphics[width=2.5in,clip=]{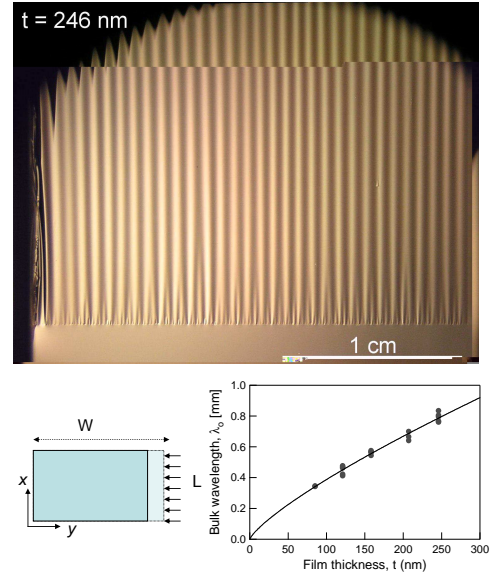}
\vspace{-3mm} \caption{\label{Image}
(A) Image of a wrinkled PS
sheet floating on the surface of water, compressed along its length
between two razor blades. (B) Sketch of geometry. (C) Bulk
wavelength of wrinkles, $\lambda=2\pi/q_o$ as a function of film
thickness, $t$.  The solid line is a fit to $t^{3/4}$, showing
agreement with the prediction of $q_{o} = \!(\rho g/B)^{1/4}$}
\vspace{-5mm}
\end{figure}

We show in Fig.\ref{Image} a controlled experimental realisation of
this situation with a polystyrene (PS) sheet of dimensions $W \times
L = 3cm \times 2cm$, and thickness $t$ floating on the surface of
water, and initially experiencing an isotropic tension $\approx\!
\gamma$, the liquid-vapour surface tension. As the sheet is
compressed by a distance $\Delta$ (such that
$\tilde{\Delta}\equiv\Delta/W\ll 1$), parallel wrinkles develop in
the bulk. The one-dimensional pattern of wrinkles in the bulk of the
sheet, characterised by a height field  $\zeta(y)$, has for small
amplitudes, an energy per unit surface area:
\begin{equation}
u = \frac{1}{2}\Big(\! B (\frac {\partial^2 \zeta}{\partial^2 y})^2
\!+\! \rho g \zeta^2 \!+\! \sigma [(\frac{\partial \zeta}{\partial
y})^2 \!-\! 2 {\tilde{\Delta}}] \Big) \label{energydensitybulk}
\end{equation}
The first two terms in $u(y)$ represent bending energy of the sheet
and the gravitational energy of the fluid, respectively, while the
third term enforces the constraint of inextensibility in the limit
of small amplitudes, with the Lagrange multiplier $\sigma$ being the
stress $\sigma_{yy}$ that must be applied in the $y$-direction at
the compressed edges. $\rho$ is the density of the fluid, and the
bending modulus is denoted by $B = Et^3/(12(1-\Lambda^{2}))$
\cite{LandauLifshitz}, where $E$ is the Young's modulus and
$\Lambda$ is the Poisson ratio. The surface tension does not appear
in the energy functional shown above due to the fact that the bulk
pattern has translational symmetry in the $\hat{x}$-direction.
However, it is important to note that the sheet still experiences a
tension $\sigma_{xx}\!\approx\!\gamma$. Minimising this energy
density leads to a pattern
\begin{equation}
 \ \zeta(y) \!=\! \frac{2}{q} \sqrt{{\tilde{\Delta}}} \sin(qy)\ \ \ \ \textbf{(a)};
 \ \sigma_{yy} \!=\! (Bq^2\!+\!\rho g/q^2)\ \ \ \  \textbf{(b)}
 \label{periodicshape}
\end{equation}
where the wavenumber $q\!=\!q_{o}\!=\!(\rho g/B)^{1/4}$. With
$q=q_{o}$, $\sigma_{yy}\!=\!-2\sqrt{B\rho g}$. As shown in
Fig.\ref{Image}, this correctly describes the scaling of the
wavelength of the wrinkles in the bulk. This scaling \cite{Cerda03}
has been experimentally tested \cite{Vella, Pocivavsek}, and more
broadly, applied in situations where the wavelength is determined by
balancing the bending energy against substrate elasticity
\cite{SIEBIMM}, capillary forces \cite{Huang2007}, and stretching
under tensile forces \cite{Cerda02}.

However, in addition to the pattern in the bulk, inspection of
Fig.\ref{Image} shows a striking phenomenon: the parallel wrinkles
of the bulk with wavenumber $q_{o}$, give way to a much finer
structure of wrinkles near the uncompressed edge of the sheet.  It
is this cascade to ever-higher wavenumber that we examine in this
Letter.

\begin{figure}
\includegraphics[width=2.5in,clip=]{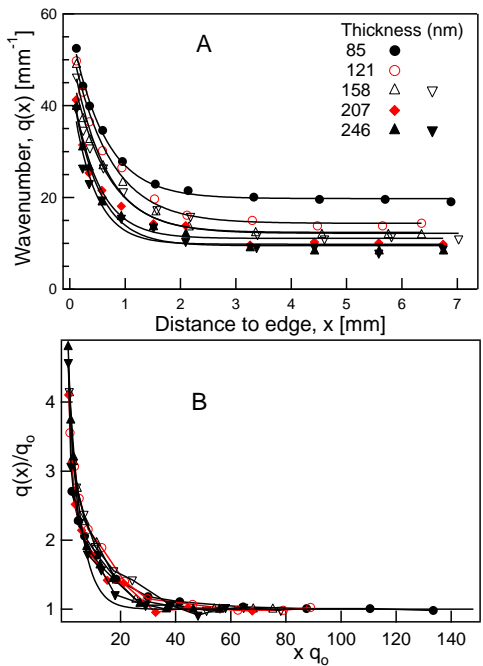}
\vspace{-3mm} \caption{\label{q_vs_x} (A) Wavenumber $q(x)$ as a
function of distance $x$ from the edge of the sheet. The lengthscale
over which the decay occurs does not change strongly with thickness
$t$ of the sheet. Exponential fits (solid lines) to $q(x)$ show
systematic deviations but yield a decay length of $1.8 \pm 0.2 mm$.
By comparison, the capillary length $l_c=2.7 mm$ (B) Scaled
wavenumber $q(x)/q_{o}$ versus the scaled distance from the edge
$xq_{o}$. Data collapse is good at small and large $xq_{o}$, but
there is poor collapse at intermediate distances indicating a
multiscale evolution. The solid line shows an exponential fit.}
\vspace{-5mm}
\end{figure}

In Fig. \ref{q_vs_x}A, we show for PS films of thickness $t$ ranging
from $85$ to $246 nm$, the increase in wavenumber $q(x)$ as a
function of distance $x$ from the uncompressed edge of the sheet.
Since the bending modulus $B\sim t^{3}$, this represents a broad
range of $B$. For all thicknesses, as the edge is approached, the
wavenumber increases to a value at the edge $q_{e}$, that is 2-5
larger than the bulk value $q_{o}$. The evolution to higher
wavenumbers occurs over approximately the same distance from the
edge: though an exponential fit shows systematic deviations, such a
fit estimates the penetration length of the edge into the bulk to be
$1.8 \pm 0.2 mm$.

Intuitively, a higher wavenumber at the edge is to be expected. The
fluid meniscus follows the contour of the edge of the sheet. To
minimise the surface energy of the air-water interface it is
therefore favourable to reduce the amplitude of the wrinkles at the
edge.  In order to achieve this while preserving inextensibility,
the wavenumber increases. This cascade to finer wrinkles must
terminate at a finite wavenumber, $q_{e}$, where the gain in surface
energy is offset by the increased energetic cost of bending.
Notwithstanding these plausible arguments, previous experiments in
this geometry \cite{Pocivavsek} did not show a marked effect at the
boundary of a film, and find that $q_{e} \!\approx\! q_{o}$. We thus
need to address some obvious issues: What are the relevant
parameters that dictate whether a cascade should be anticipated?
When a cascade is observed, what governs the amplification of the
bulk wavelength, and the length over which the cascade occurs?

In order to understand why wavenumber amplification occurs in our
experiment, we estimate the energy cost of a wavenumber $q_{e}$ at
the edge. The capillary energy associated with the meniscus is
$U_{cap}=2\gamma ({\tilde{\Delta}}/q_{e}^{2})\sqrt{(\rho
g/\gamma)+q_{e}^{2}}$, which as sketched in Fig. \ref{sketch}A,  is
a decreasing function of $q_{e}$. To compare $U_{cap}$, which is an
energy per unit length, to the energy cost per unit area of the
affected part of the sheet (Fig. \ref{sketch}B), we require an
understanding of the lengthscale over which the pattern at the edge
penetrates into the sheet and matches the bulk pattern.
\begin{figure}
\includegraphics[width=2.5in,clip=]{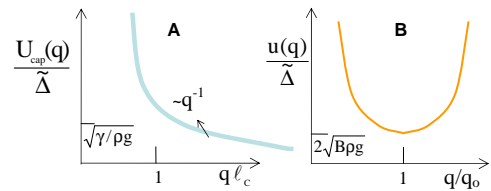}
\vspace{-1mm} \caption{\label{sketch} (A) The energy per length of
the meniscus, $U_{cap}(q_{e})=2\gamma({\tilde{\Delta}}/q_{e}^{2})
\sqrt{(\rho g/\gamma)+q_{e}^{2}}$. (B) Energy per area of a sheet
with wavenumber $q$, $u(q)=2(\tilde{\Delta}/q^2)B(q^4+q_{o}^4)$.}
\vspace{-5mm}
\end{figure}

This leads us to consider the effect on the energy of the sheet of
imposing a wavenumber $q_e$ at its edge. The effect of broken
translational symmetry in the $\hat{x}$-direction is incorporated by
modifying Eq. \ref {energydensitybulk} to:
\begin{equation}
u = \frac{1}{2}\Big(\! B (\nabla^2 \! \zeta)^2 \!+\! \rho g \zeta^2
\!+\gamma (\frac{\partial \zeta}{\partial x})^2+\! \sigma(x)
[(\frac{\partial \zeta}{\partial y})^2 \!-\! 2 {\tilde{\Delta}}]
\Big). \label{energydensitygeneral}
\end{equation}
The new, third, term, $u_T$ quantifies the energy cost of the
deformations in the $\hat{x}$-direction under the tension $\gamma$
of the fluid tugging at the uncompressed edge of the sheet. When the
compressive force in the y-direction is much smaller than the
tensile force in the x-direction, $\varepsilon \equiv \gamma/\sigma
= \gamma/\sqrt {\rho g B}\ll1$, the minimal energy profile is
determined by balancing bending and tensile forces [see
Supplementary Information]. We thus develop a length scale, $l_{p}$,
that will govern gradients in the x-direction. From $Bq_{o}^{4} \sim
\gamma l_{p}^{-2}$, we obtain $l_{p}\approx \sqrt {\gamma/ \rho g}$,
which is the capillary length, $l_c$. This is consistent both with
the magnitude of the typical penetration length found in Fig.
\ref{q_vs_x}, and with the insensitivity of this length scale to the
thickness $t$. Furthermore, the stress ratio $\varepsilon$ ranges
between $6\times10^{-4}$ and $3\times10^{-3}$ in our experiments,
thus validating the regime assumed in the foregoing argument.

A self-consistent argument for the strength of the wavelength
amplification may be constructed by assuming that a single
penetration length, $l_{p}$, governs the penetration of the edge
wavenumber into the bulk; an estimate of the bending energy cost
near the edge of the sheet is then obtained as $U_{edge}\sim
l_{p}u(q_{e})\sim Bq_{e}^{4} (\Delta/q_{e}^{2})$. Setting the edge
and meniscus energies to be comparable, $U_{edge} \sim U_{cap}$ we
obtain $q_{e} \sim (\varepsilon)^{-1/3} \sqrt{(\gamma/B)}$. This
argument shows that for $\epsilon \ll1$ (as in our experiments) the
wavenumber amplification is a large, non-perturbative effect on the
bulk pattern. However, the estimate $U_{edge} \sim q_{e}^2 $ derived
above is wrong as it relies on the oversimplified picture of a
single penetration length, unlike the cascade observed in Fig. 2
which indicates an elastic instability that leads to a
symmetry-breaking sequence of period fissioning into intermediate
wavenumbers (and correspondingly a sequence of penetration lengths).
A full solution of the nonlinear problem \cite{Benny2008} reveals
that this instability results from a logarithmic dependence of
$U_{edge} \sim \log q_e$, which lowers the energetic cost for
sufficiently large values of $q_e$.


Assuming a pattern that is the superposition only of two wavenumbers
$q_0$ and $q_e$, the penetration depth $l_p$ exhibits a strong
dependence on $\epsilon$, ranging from $l_c$ for $\epsilon \ll 1$ to
$q_0^{-1}$ for $\epsilon\gg 1$ (see Supplementary material). The
latter limit, together with an energetic estimate similar to the one
presented above, shows that for $\epsilon \gg 1$ the edge effect is
only a small perturbation to the bulk pattern \cite{Pocivavsek}. For
$\epsilon \ll 1$ when there are wavenumbers $q$ intermediate between
$ q_0$and $q_e$, one can show that the length $\l_p(q)$ of the
transition zone between wavenumbers $q$ and $q_e$, strongly depends
on $q$, ranging from $l_c$ for $q \to q_0$ to $q_e^{-1}$ for $q \to
q_e$. This is because a periodic wrinkling pattern with wave number
$q>q_0$ requires enhancement of the compression $\sigma(x)$ towards
the edge (thus increasing $\epsilon$ close to the edge). That the
description in terms of a single penetration length $\l_p$ is
simplified can be seen from Fig. 2B, where we show the scaled
wavenumber $q(x)/q_0$ vs. the scaled distance from the edge $x q_0$.
As anticipated, this scaling yields good data collapse far from the
edge, and even close to the edge. However, data does not collapse in
the intermediate region, indicating a wave-dependent penetration
length $\l_p(q)$.

Thus far, our estimates of the overall lengthscales of wavenumber
amplification have shed no light on the nature of the
cascade itself.  
The classic example of an elastic cascade was predicted by Pomeau
and Rica \cite{PomeauRica} for the so-called "curtain geometry" of a
tension-free sheet rippled under a compressive force but constrained
to be flat at one edge. They showed that the matching of the ripples
to the flat edge ($q_e\to\infty$) was achieved by an infinite
hierarchy of branching events, in which each wrinkle branches into a
succession of sharp folds with flat faces.

\begin{figure}
\includegraphics[width=2.5in,clip=]{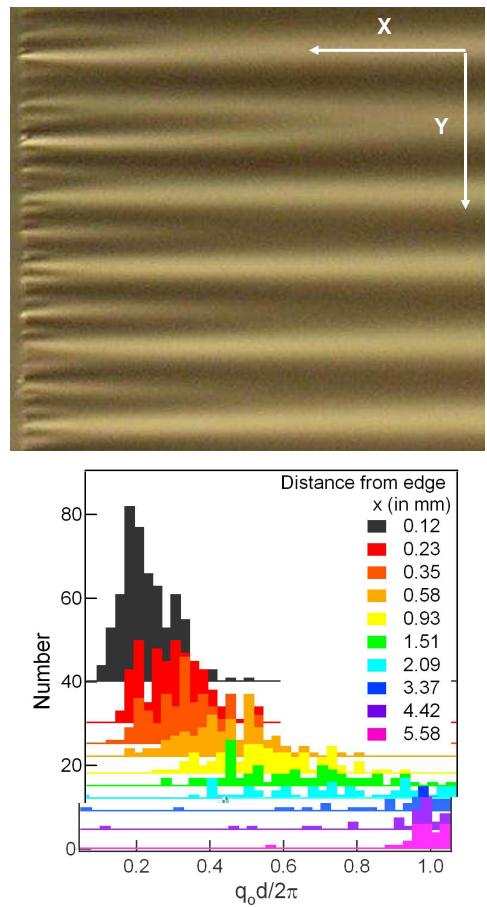}
\vspace{-3mm} \caption{\label{sketch} (A) A magnified image of the
cascade.  (B)  At each value of $x$, a histogram of the scaled
separation, $q_{o}d/(2\pi)$, between crests, for several values of
distance $x$ from the edge. Data were collected from two films with
$t=246nm$. The separations $d$, are determined from the locations of
maxima of the intensity in the $\hat{y}$ direction.} \vspace{-5mm}
\end{figure}

One superficial difference between what we observe and the
Pomeau-Rica cascade is that our cascade terminates at a finite
wavenumber, and therefore passes through only a few generations. The
more profound difference stems from the fact that our sheets
experience a tension along the uncompressed direction. Any deviation
from a one-dimensional wrinkling pattern imposes curvature in both
directions; this Gaussian curvature generates in-plane stretching
energy controlled by a modulus $Y = Et$. In the tensionless
Pomeau-Rica scenario, the dominant contribution to the strain energy
is the anharmonic energy density $u_G \sim Y \zeta_x^2 \zeta_y^2 $,
whose minimisation leads to localised Gaussian curvature along a
sequence of sharp ridges \cite{Witten07}. However, the consequence
of the applied tension is that the focusing of Gaussian curvature
does not fully relieve the strain energy at other points: as noted
in Eq. \ref{energydensitygeneral}, the tension term $u_{T}$
penalises slope, and is non-zero even on flat facets where the
gaussian curvature vanishes. This mechanism thus favours a smooth
reduction of the amplitude, which is naturally accomplished by the
superposition of a finite number of Fourier modes with distinct
wavenumbers. A transition between the Pomeau-Rica cascade and a
smooth pattern is expected if $u_T \approx u_G$, which implies [see
Supplementary Material] $\tilde{\Delta} <T/Y$. Together with the
threshold condition for wrinkling $\tilde{\Delta} > \sigma_0/Y$ we
see that a necessary condition for a smooth pattern is that the
tension be large enough such that $\varepsilon \lesssim 1$. The
characteristic values of $\varepsilon$ in our experiments are well
within this regime indicating a novel, smooth hierarchy, markedly
different from the Pomeau-Rica stress-focusing cascade.

A closer look at the cascade, as shown in the magnified view of Fig.
4A, supports this new scenario of a smooth mechanism in which larger
amplitudes of higher wave-number Fourier components are smoothly
mixed in as one approaches the edge.  In Fig. 4B we present a more
quantitative measure of the smoothness of the cascade. At a given
distance $x$ from the uncompressed edge we determine from the image,
the separations $d$, between the crests of the wrinkles.  At each
value of $x$, we show a histogram of $q_{o}d/(2\pi)$, the normalised
separation between wrinkles. Far away from the edge, the separations
are all concentrated at $q_{o}d/(2\pi)=1$.  As expected, closer to
the edge, more crests are formed, and at smaller values of $d$.
Importantly, none of the histograms show significant weight near
$d=0$.  In a scenario where wrinkles divide by localised branching,
one might expect a preponderance of small values of $d$ just after a
branch-point, between sibling branches of the same parent wrinkle.
That appears not to be the case in Fig. 4, with separations flowing
smoothly to a mixture of higher Fourier components.

The three forces operative in this problem -- gravity, bending
energies, and surface energies -- can be combined in pairs to yield
three distinct length-scales: $(\rho g/B)^{1/4}$,
$\!(\gamma/B)^{1/2}$ and $l _{c}=\!(\gamma/\rho g)^{1/2}$.  The
first of these is $q_{o}$, the second, an elastocapillary length
\cite{Bico2004} which controls $q_{e}$, and the third is the
capillary length, which determines the length of the cascade.
However, we emphasise that the stress ratio $\varepsilon =
\sigma/\gamma$ dictates the overall morphology of the pattern in our
experiments. Surface tension plays a dual role in our experiments,
determining both the energy of the fluid meniscus, $U_{cap}$, as
well as the tension applied in the uncompressed direction. In
principle, these are different effects that could be independently
tuned. Increasing the capillary energy cost of the edge can tune a
transition from the regime of our experiment to that of stiffer
sheets, in which the effect of the edge is small. On the other hand,
decreasing the applied tension could drive a transition from the
smooth, energy-delocalised cascades we observe, to a regime of
localised branching \cite{PomeauRica} with energy-focusing. Thus,
our observations open the way to the exploration of a rich phase
diagram \cite{Benny2008} of both branched and smooth structures.

\textbf{Materials and Methods}:  The films were prepared from
solutions of polystyrene (PS; atactic, number-average molecular
weight $M_{n} = 121K$, weight-average molecular weight $M_{w} = 1.05
M_{n}$, radius of gyration $R_{g} ~ 10 nm$) in toluene, spin-coated
on to glass substrates. The film thickness $t$ was varied by
changing the concentration of the solution and the spin rate, and
was measured by x-ray reflectivity using a Panalytical X-Pert x-ray
diffractometer. The $CuK\alpha$ radiation from the x-ray source
(wavelength $\lambda = 0.154 nm$) is coupled to a parabolic, graded
multilayer mirror assembly that produces a low-divergence beam of
x-rays incident on the PS film. Observation of the Kiessig fringes
in specular reflection with a two-circle goniometer  yielded the
film $t$ with a precision of $\pm 0.5 nm$.

A rectangle of dimensions $L \times W = 2 cm \times 3 cm$ was
scribed onto the film with a sharp edge. When the substrate was
dipped into a petri dish of distilled, deionized water, a
rectangular piece of the PS film detached from the substrate.
Because PS is hydrophobic, the film remained at the air-water
interface.  In order to prepare extremely well-defined edges, in a
few selected cases, we spin-coated the film on a silicon wafer with
an oxidised surface layer. Rather than scribing the film, the
silicon substrate was fractured along a crystalline plane, and the
PS film was lifted off with HF solution.

\textbf{Acknowledgments} We acknowledge support from the Center for
University of Massachusetts Industry Cooperative Research Program
(JH); the NSF–supported MRSEC on Polymers at the University of
Massachusetts; the U.S. DOE, through DE-FG-0296 ER45612 (TPR); and
the NSF through NSF-DMR 0606216 (NM) and NSF-CBET-0609107 (TPR, NM).
We thank B. Roman, R. Kamien, and D.R. Nelson for useful
conversations.

\textbf{Author contributions} Experiment design: JH, TPR, NM; Data:
JH; Analysis: JH, NM with ideas developed with BD; Theory: BD, CS;
Manuscript preparation: NM, BD with input from TPR, CS.

\addcontentsline{toc}{chapter}{\rm {\bf BIBLIOGRAPHY}  . . . . . . .
. . . . . . . . . . . . . . . . . . . . . . . . . . . . . . . .}

\end{document}